\documentclass[runningheads]{llncs}
\usepackage{array}
\usepackage[numbers]{natbib}

\usepackage[T1]{fontenc}
\usepackage{multirow}
\usepackage{booktabs}
\usepackage{blindtext}
\usepackage{graphicx}
\begin{document}
\title{A Participatory Strategy for AI Ethics in Education and Rehabilitation grounded in the Capability Approach}
\titlerunning{Participatory Strategy}
%
\author{Valeria Cesaroni\inst{1}~\orcidID{0000-0001-9881-1597} \and
Eleonora Pasqua\inst{4,5}~\orcidID{0000-0002-7153-6094} 
\and
Piercosma Bisconti\inst{3}~\orcidID{0000-0001-8052-0142} \and 
Martina Galletti\inst{2,3}~\orcidID{0009-0002-2079-8999}}
\authorrunning{V. Cesaroni, E. Pasqua, P. Bisconti and M. Galletti}
%
\institute{University of Perugia, Italy \and 
Sony Computer Sciences Laboratories - Paris, France \and 
Sapienza University of Rome, Dipartimento di Ingegneria Informatica, Automatica e Gestionale "A. Ruberti" \and 
Centro Ricerca e Cura di Roma, \and 
Sapienza University of Rome, Dipartimento di Organi di Senso, 
\email{valeria.cesaroni@dottorandi.unipg.it} \\
\email{martina.galletti@sony.com}\\
%
}
\maketitle              
\vspace{-18pt}
\begin{abstract}
AI-based technologies have significant potential to enhance inclusive education and clinical-rehabilitative contexts for children with Special Educational Needs and Disabilities. AI can enhance learning experiences, empower students, and support both teachers and rehabilitators. However, their usage presents challenges that require a systemic-ecological vision, ethical considerations, and participatory research. Therefore, research and technological development must be rooted in a strong ethical-theoretical framework. The Capability Approach - a theoretical model of disability, human vulnerability, and inclusion - offers a more relevant perspective on functionality, effectiveness, and technological adequacy in inclusive learning environments. In this paper, we propose a participatory research strategy with different stakeholders through a case study on the ARTIS Project, which develops an AI-enriched interface to support children with text comprehension difficulties. Our research strategy integrates ethical, educational, clinical, and technological expertise in designing and implementing AI-based technologies for children's learning environments through focus groups and collaborative design sessions. We believe that this holistic approach to AI adoption in education can help bridge the gap between technological innovation and ethical responsibility.
\vspace{-10pt}
\end{abstract}
\vspace{-28pt}
\section{Introduction}
\vspace{-8pt}
The integration of artificial intelligence (AI) in educational and clinical contexts necessitates a multidimensional approach that considers cognitive, relational, and ethical dimensions. First, the literature underscores that artificial intelligence, by directly influencing cognitive and relational structures~\citep{chiriatti2024case} can play a crucial role in the scaffolding function. This, in turn, plays a pivotal role in proximal development, significantly shaping subjectification, qualification, and socialization processes~\citep{tuomi2022artificial}. Second,  an inclusive perspective requires an ecological-systemic approach to the technology design that can take into account the interaction between individuals, technologies, and their broader social environments~\citep{cesaroni2024towards}. For these reasons, the development and the implementation of such technologies must be anchored within an ethical framework which is appropriate and grounded in a theoretical model suitable for childhood, disability, and human vulnerability~\citep{tuomi2023framework}. Moreover, it is essential to design the development of the technology by engaging diverse stakeholders, each bringing distinct epistemic and methodological perspectives. 


This study makes several key contributions. Drawing inspiration from the ARTIS Project~\citep{galletti2023reading, galletti2024artis}, which develops an AI-based interface to rehabilitate children with text comprehension difficulties. The study contributes to:  (I) Propose a theoretical framework based on the capability approach;  (II) Introduce a co-creation strategy applicable to various contexts;  (III) Present implementation results using ARTIS as a case study. The study illustrates a research and development pathway linking inclusion, technological innovation and critical assessment of personal capabilities, proposing a holistic and participatory model for ethical and inclusive AI development aimed at supporting vulnerable learners.

\vspace{-18pt}
\section{Related Work}
\vspace{-5pt}
The field of AI in Education (AIEd) has grown alongside developments in big data and adaptive technologies, with AI typically supporting students, teachers, or institutions~\citep{holmes2022state}. While widely adopted in mainstream education for administration and personalized learning~\citep{pagliara2024integration}, its role in inclusive and special education remains limited and often instrumental~\citep{barua2022artificial}. Scholars have criticized the lack of ethical and pedagogical consideration in AI integration~\citep{mouta2024uncovering} noting that many tools are not designed with the needs of learners with disabilities in mind~\citep{holmes2022state}.

Recent research has moved beyond AI’s technical functions to highlight its ethical and social implications in education~\citep{holmes2023ethics}. AI ethics research often remains narrowly technical—focused on fairness, bias, and transparency—while neglecting deeper pedagogical and inclusive concerns~\citep{cesaroni2024inclusive,cesaroni2024towards}. This functionalist approach sidelines critical, theory-driven perspectives and rarely involves educators or specialists in early design stages~\citep{zawacki2019systematic,seon2024ai}. The lack of an ecological and systemic outlook results in AI tools misaligned with pedagogical values and classroom realities, especially in inclusive education contexts.

Finally, AI has demonstrated promise in diagnosing neurodevelopmental disorders through machine learning analysis of MRI ~\citep{cao2023machine} and EEG ~\citep{ahire2023comprehensive} data, providing effective screening support~\citep{khan2018machine}. Despite benefits for clinical decision-making ~\citep{merzon2022eye,vadala2024uso} and personalized learning~\citep{reiss}, AI's role in structured therapeutic interventions and long-term rehabilitation remains underexplored. Research on AI in structured therapeutic interventions is still limited~\citep{bhardwaj2024transforming}, and few programs provide adequate training for educators and clinicians.

This study addresses the lack of AI-driven rehabilitation tailored to diverse cognitive needs and the shortage of training for practitioners by positioning AI as a core component of clinical and educational interventions that support cognitive and linguistic development. The ARTIS project~\citep{galletti2023reading,galletti2024artis} exemplifies a research approach based on an ethical framework of human development and provides an example of the context-sensitive use of technology. Based on van Dijk's neuropsycholinguistic model~\citep{kintsch1978toward,van1983strategies}, and developed with speech therapists and tested in public and clinical settings~\citep{galletti2024artis}, ARTIS aims to support the rehabilitation process. 
\vspace{-15pt}
\section{The Theoretical Framework}
\vspace{-10pt}
Existing ethical frameworks for AI, often focused on risk mitigation (e.g. fairness, bias, privacy), prove inadequate in educational and rehabilitative contexts involving children. These models tend to be object-oriented and contractualist, assuming autonomous, equal agents~\citep{tuomi2023framework}. Such assumptions neglect the fundamental conditions of vulnerability, dependence, and asymmetry that characterize childhood and disability~\citep{tuomi2023framework}.
We argue for a shift toward a holistic perspective that sees technology as a sociotechnical mediator rather than a neutral tool~\citep{latour2005reassembling}. This requires an ethical model grounded in vulnerability, interdependence, and the complex interplay between individuals, technology, and their environments. AI, especially in childhood contexts, challenges instrumentalist perspectives and demands a rethinking of inclusion through a more relational and systemic lens.
To address this gap, we adopt the Capability Approach (CA) as a foundational framework. Originating with Amartya Sen~\citep{sen2000development} and expanded by Martha Nussbaum~\citep{nussbaum2007frontiers}, the CA redefines justice not in terms of equal resources, but of real freedoms - what individuals are effectively able to do and be. It emphasizes conversion factors - individual (e.g. impairments), social (e.g. norms, expectations), and environmental (e.g. infrastructures) - that shape how resources like technology translate into meaningful opportunities (capabilities and functionings)~\citep{coeckelbergh2012learned}.
In this light, AI must be evaluated not just on its technical performance but on how it shapes learning, inclusion, and well-being. It calls for research questions attentive to human diversity, contextual factors, and the mediating role of technology in subject formation.
With this approach, we can interpret the interplay between technology, context, and users’ situated conditions in a more nuanced way, enabling a deeper, more critical understanding of the role technology plays in inclusive processes.
Especially relevant to AI - defined as a third-level technology~\citep{floridi2014fourth} for its pervasive cognitive and hermeneutic role - the CA supports a systemic, hybridized view of educational processes.

In the ARTIS project, the Capability Approach has shaped the formulation of key research questions that guide both design and evaluation. These include: whether the technology is developmentally appropriate for its intended users; what user model underpins the system's design choices; and which capabilities are promoted or, conversely, constrained by the intervention. Additional considerations include how the processes of subjectification, qualification, and socialization are influenced by the technology, and what forms of human–AI interaction best support learner agency. In sum, the CA offers a rich ethical and epistemological lens to reframe inclusion and justice in the age of AI, moving beyond risk management toward a vision of relational, situated, and developmental justice. 
\vspace{-15pt}
\section{The Co-Creation Framework: Principles and Practical Applications}
\vspace{-5pt}
\subsection{General Principles}
The ARTIS project has developed a participatory research strategy for interface design that combines the Capability Approach with Value Sensitive Design~\citep{friedman2019value,galletti2024artis}. This ensures AI-driven educational tools align with real-world needs and pedagogical goals by involving educators, clinicians, and developers from the start. Unlike traditional methods that seek input after development, ARTIS emphasizes early, interdisciplinary co-creation. This approach, adaptable beyond ARTIS, serves as a model for ethical, effective AI in education and rehabilitation. In this section, we introduce CORE – the Co-Creation framework for Optimized, Responsible, and Ethical AI Development – our second technical contribution, detailing its structure, methodology, and principles.

CORE unfolds in three phases to align diverse perspectives across tech, clinical, and educational fields. Phase one builds a proof of concept. Phase two refines the interface through ethical, pedagogical, and functional insights grounded in focus groups and interviews with developers and clinicians. Phase three evaluates the tool’s impact and ethics in real-world settings, using the Capability Approach to ensure meaningful, responsible learning outcomes. To provide a clear overview, Table~\ref{tab:framework} presents the CORE model as a structured framework for guiding AI development in educational and therapeutic contexts. It highlights six key dimensions that support ethical, pedagogical, and human-centered design, implementation, and evaluation. Each dimension includes key points and guiding questions to prompt critical reflection and decision-making. Results from the ARTIS project are discussed in detail in Section~\ref{practical-app}.

The CORE framework is grounded in the following guiding principles:
\vspace{-5pt}
\begin{enumerate}
    \item \textbf{Child-Centered Design} – Defines objectives, age-appropriateness, and developmental alignment while mapping pedagogical models and impacted capabilities.
    \item \textbf{Human Interaction and Context} – Specifies use cases and aligns expectations across educational and clinical environments.
    \item \textbf{Participatory Engagement} – Involves stakeholders (e.g., teachers, families, clinicians, children), addresses value tensions, and supports iterative co-design.
    \item \textbf{Agency} – Examines the balance between automation and human oversight, autonomy, privacy and its impact on users’ self-efficacy and working conditions.
    \item \textbf{AI Quality and Robustness} – Focuses on data quality, fairness, transparency, and privacy, especially in sensitive child-focused applications.
    \item \textbf{Experimental Validation} – Assesses functionality, acceptance, and risks, ensuring alignment with user values and identifying barriers and enablers.
\end{enumerate}
\vspace{-8pt}
We introduce Table~\ref{tab:framework}, as a conceptual and practical resource, supporting AI development that emphasizes ethical integrity, user needs, and contextual relevance in educational and clinical environments. 
\vspace{-10pt}
\begin{table}[h!]
    \centering
    \renewcommand{\arraystretch}{1.2} 
    \setlength{\tabcolsep}{6pt} 
    \resizebox{0.95\textwidth}{!}{
    \begin{tabular}{>{\centering\arraybackslash}m{3.3cm} 
                    >{\raggedright\arraybackslash}m{5.5cm} 
                    >{\raggedright\arraybackslash}m{7cm}} 
        \toprule
        \toprule
        \textbf{Conceptual Dimension} & \textbf{Key Considerations} & \textbf{Guiding Questions} \\
        \midrule
        \textbf{Children-Centered Framework} &  
        \begin{itemize} \setlength{\itemsep}{2pt}
            \item Clear definition of technology’s objectives
            \item Alignment with developmental stages
            \item Identification of impacted capabilities
            \item Explicit pedagogical models
        \end{itemize} &  
        \begin{itemize} \setlength{\itemsep}{2pt}
            \item What are the intended learning outcomes?
            \item How does it support developmental stages?
            \item What cognitive and social skills are targeted?
            \item Which pedagogical models underpin the design?
        \end{itemize} \\
        \midrule
        \textbf{Human Interaction and Setting} &  
        \begin{itemize} \setlength{\itemsep}{2pt}
            \item User experience and accessibility
            \item Clear interaction models
        \end{itemize} &  
        \begin{itemize} \setlength{\itemsep}{2pt}
            \item What are the primary use cases?
            \item How can stakeholder alignment be achieved?
        \end{itemize} \\
        \midrule
        \textbf{Democratic Engagement} &  
        \begin{itemize} \setlength{\itemsep}{2pt}
            \item Inclusive stakeholder involvement
            \item Addressing value conflicts
            \item Systematic feedback integration
        \end{itemize} &  
        \begin{itemize} \setlength{\itemsep}{2pt}
            \item How can educators, parents, and therapists contribute?
            \item What mechanisms ensure inclusive evaluation?
        \end{itemize} \\
        \midrule
        \textbf{Agency} &  
        \begin{itemize} \setlength{\itemsep}{2pt}
            \item Balance between automation and human oversight
            \item Ensuring autonomy and privacy
        \end{itemize} &  
        \begin{itemize} \setlength{\itemsep}{2pt}
            \item Is the technology overly intrusive?
            \item How does it influence learning and interaction?
            \item What are its implications for educators and clinicians?
        \end{itemize} \\
        \midrule
        \textbf{AI Quality and Robustness} &  
        \begin{itemize} \setlength{\itemsep}{2pt}
            \item Data quality and bias mitigation
            \item Algorithmic fairness and transparency
            \item Protection of user privacy
        \end{itemize} &  
        \begin{itemize} \setlength{\itemsep}{2pt}
            \item How is data preprocessed and validated?
            \item What measures ensure algorithmic fairness?
            \item How is children’s data privacy safeguarded?
        \end{itemize} \\
        \midrule
        \textbf{Experimental Validation} &  
        \begin{itemize} \setlength{\itemsep}{2pt}
            \item Technology acceptance assessment
            \item Rigorous testing and risk analysis
        \end{itemize} &  
        \begin{itemize} \setlength{\itemsep}{2pt}
            \item What social acceptance frameworks are considered?
            \item What barriers may hinder stakeholder adoption?
            \item Has a comprehensive risk mitigation plan been implemented?
        \end{itemize} \\
        \bottomrule
        \bottomrule
    \end{tabular}%
    }
    \vspace{10pt}
    \caption{This table outlines a structured framework for developing AI technologies. The table identifies six fundamental aspects essential for evaluating AI applications in child-focused contexts, highlights critical factors within each dimension, and poses reflective questions to ensure ethical and developmental alignment.}
    \label{tab:framework}
    \vspace{-20pt}
\end{table}
\vspace{-25pt}
\subsection{Practical Applications}\label{practical-app}
\vspace{-5pt}
The ARTIS project is currently in Phase Two of the CORE Framework, focusing on qualitative analysis through two focus groups and rounds of semi-structured interviews with two key practitioner groups: AI engineers/developers and clinical professionals. While the former manages software design and implementation, the latter evaluates its rehabilitative functionality. Insights from these sessions highlighted the need for a more holistic, ethically grounded development strategy, emphasizing alignment among the project's technical, clinical, and educational pillars. Initial studies involved the “AI Group” ($n = 6$), comprising full-stack developers and engineers who assessed data quality, pictogram and vocabulary appropriateness, human oversight, and human-computer interaction. Subsequently, the “Clinicians Group” ($n = 6$) from the Research and Care Center in Rome (CRC) evaluated the technology’s rehabilitative relevance, inclusivity, user expectations, and ethical implications in clinical contexts.

One of the most extensively analyzed aspects in the discussion with clinicians was the modalities of human oversight and interaction, both in the direct engagement between clinicians and the software and within the broader clinical-rehabilitative setting.  The primary objective was to understand the needs and challenges associated with integrating technology into rehabilitation contexts and to evaluate the theoretical alignment between the AI Group and the Clinician Group. Within the framework of the Capability Approach, achieving an optimal balance of human oversight is critical.  Specifically, it is essential to mediate between (a) the degree of automation, which ensures that the technology remains unobtrusive and facilitates seamless interaction, and (b) the necessity for clinicians to retain control and the ability to intervene when using the interface.  While excessive automation would entail risks associated with the loss of adult agency in the use of technology, an excessive need for intervention could potentially disrupt the fluidity of the setting, leading to a reversal of the technology's intended function. The goal is for the technology to serve as a scaffolding tool, subtly supporting the interaction between the child and the educator or rehabilitator rather than becoming the dominant element in the learning process. If human oversight is not appropriately balanced, there is a risk that the interaction will shift from a child-centered learning dynamic to one where managing the technology takes precedence. Another key finding was the need to align theoretical and practical models between groups, particularly in user experience, interface design, and interaction mechanisms. In response to the focus group, a second software prototype was developed to better integrate technical and therapeutic requirements.
\vspace{-10pt}
\section{Conclusion \& Future Work}
\vspace{-5pt}
This study underscores the importance of a multidimensional, ethical, and inclusive approach to integrating AI in education, emphasizing its potential to foster cognitive and relational development in alignment with educational goals such as subjectification, qualification, and socialization. Adopting a systemic-ecological perspective, it considers the dynamic interplay between individuals, technology, and environments—particularly for vulnerable populations. Using an AI-powered reading comprehension interface as a case study, the research contributes by: (I) establishing a capability-based theoretical framework for AI development, (II) implementing a stakeholder-driven co-creation strategy, and (III) demonstrating it through a case study. Future work will develop the framework into a systematic methodology for evaluating ethical issues in child-centered AI. This methodology will offer a systematic approach to assessing ethical considerations across various AI applications. 
\bibliographystyle{splncs04}

\end{document}